\documentclass[referee]{aa} 

\usepackage{graphicx}
\usepackage{txfonts}
\usepackage{epsfig}
\usepackage{tabularx}
\usepackage{supertabular}

\begin{document}
\title{Synthetic horizontal branch morphology for different
    metallicities and ages under tidally enhanced  stellar wind}

\author{Z. Lei
  \inst{1,2,3}
    \and F. Zhang
  \inst{1,2}
      \and
  H. Ge
      \inst{1,2}
  \and
  Z. Han \inst{1,2}
}


\institute{National Astronomical Observatories/Yunnan Observatory,
  the Chinese Academy of Sciences, KunMing 650011, China\\ 
  \email{lzx2008@ynao.ac.cn}
  \and
  Key Laboratory for the Structure and Evolution of Celestial Objects, the Chinese Academy of Sciences, Kunming 650011, China 
  \and
  University of the Chinese Academy of Sciences, Beijing 100049, China}

\date{Received ; accepted}


\abstract
{It is believed that some other parameters,
except for metallicity (the first parameter),
are needed to explain the horizontal branch (HB) morphology
of globular clusters (GCs).  Furthermore,
these parameters are considered to be correlated with
the mass loss of the red giant branch (RGB) stars. Unfortunately, the
physics of mass loss on the RGB is poorly understood at present.
In our previous work, we proposed that 
tidally enhanced stellar wind during binary evolution may affect the
HB morphology by enhancing the mass loss of the 
red giant primary and that we can reproduce the basic morphology of HB in GCs.   }
{We did not consider the effects of other
important parameters (e.g., metallicity and age) in
our final results there. 
As a further study, we now
investigate the effects of metallicity and age  
on HB morphology by  considering
tidally enhanced stellar winds during binary evolution.  }
{We incorporated the tidally enhanced-stellar-wind model of Tout \& Eggleton into Eggleton's stellar evolution code to study the binary evolution.
A group of binary system samples were generated by Monte Carlo
simulations. The position of each sample star
in a color-magnitude diagram was obtained
by transforming temperature and luminosity into
$B-V$ color and absolute magnitude.
To study the effects of metallicity and age on our final
results, we conducted two sets of model calculations:
(i) for a fixed age, we used three metallicities,
namely $Z$=0.0001, 0.001 and 0.02.
(ii) for a fixed metallicity, $Z=0.001$,
we used five ages in our model calculations:
14, 13, 12, 10, 7 Gyr.}
{We found that HB morphology of GCs
becomes bluer with decreasing metallicity,
and old GCs present bluer HB morphology than young ones.
These results are consistent with previous work.
Although the envelope-mass distributions of zero-age
HB stars produced by 
tidally enhanced stellar wind are similar
for different metallicities,
the synthetic HB under tidally enhanced
stellar wind for $Z$=0.02 presented a distinct gap 
between red and blue HB. However, this feature 
was not seen clearly in the synthetic HB for $Z$=0.001 and 0.0001. 
We also found that binary fractions may 
make HB morphology become bluer, and 
we discuss the 
results with recent observations.} 
{}


\titlerunning{Synthetic HB morphology in GCs}
\authorrunning{Z. Lei et al. }

\maketitle

%

\section{Introduction} \label{1. Introduction}
The color distribution of horizontal branch (HB) stars in
the color-magnitude diagram (CMD) of globular clusters (GCs),
which is defined as HB
morphology,  is a longstanding problem
in stellar evolution.  Although the most important
factor affecting HB morphology is identified as the
metallicity (called the first parameter, Sandage \& Wallerstein 1960),
other parameters are also believed to play an important
role in this problem. These parameters are called the
second parameter (2P).  However,
the nature of the 2P is still unknown.
Many 2P candidates have been proposed in the past five decades 
(for a recent review see Catelan 2009; also see Lei et al. 2013; here 
after Paper I), 
but none of them can explain the whole HB morphologies  
in GCs. 
Recently, some researchers have 
suggested that, except for metallicity, age may be the main global
parameter affecting HB morphology in GCs, while some other parameters
are needed as the third parameter, such as helium abundance and 
central density (Gratton et al. 2010; Dotter et al. 2010).
Moreover, the second parameter problem is considered to be
correlated with multiple populations discovered in GCs
(Piotto et al. 2007, 2012; Carretta et al. 2010; 
Gratton, Carretta \& Bragaglia, 2012 ).

Mass loss on the red giant branch (RGB) is a very important process in 
understanding the 2P problem in GCs 
(Catelan 2009), 
since the color distribution of HB stars in CMD depends on
their envelope masses, while the envelope mass is determined by
the mass loss on the RGB.
Unfortunately, the physics of mass loss on the RGB is poorly known at present
(Willson 2000; Dupree et al. 2009).
Thus, in order to reproduce 
the color distribution of HB stars in the CMD of GCs, most of the 2P candidates 
need to assume a
mass-loss dispersion on the RGB (e.g., Gaussian distribution of
mass with a mass dispersion for
HB stars; Lee, Yoon \& Lee 2000; Lee et al. 1994; D'Antona \& Caloi, 2004;
Percival \& Salaris 2011).
However, the assumption of mass-loss dispersion is
arbitrary and has no any physical justification.

In our previous work (see Paper I), we proposed that
tidally enhanced stellar wind during binary evolution
may affect the HB morphology in GCs by enhancing the
mass loss of the red giant primary.  In a red giant binary
system, the
stellar wind of the red giant primary may be
tidally enhanced by the secondary.
The exact mass loss of the primary stars on the RGB depends on the
initial binary orbital period. Therefore
different initial binary orbital periods
lead to different mass loss of the primary on the RGB.
After the ignition of helium in their cores,
the primary stars are located in different positions of the
HB.  In Paper I, under  consideration of
tidally enhanced stellar wind and without
any additional assumptions of mass-loss dispersion
on the RGB, we can reproduce the
basic HB morphology of GCs, in which red, blue, and
extreme HB are all populated.
However, we do not know exactly how
other important parameters (e.g., metallicity and age) 
affect our results obtained in Paper I. 
As further study, 
in this paper, using 
the same method as described in
Paper I, we study the effects on HB morphology
for different metallicities and ages 
by considering the tidally enhanced stellar
wind during binary evolution. 

The structure of this paper is as follows. In Section~2, we introduce
the method and code used in this paper. 
Our results  are given in Section~3.  A discussion and conclusions are given in Section~4 and Section~5, respectively.

\section{Method and code}
As described in Paper I, in this work, we also incorporate
tidally enhanced stellar wind during binary
evolution (see Sect~2.2) into
Eggleton's stellar evolution code (see Sect~2.1) to
calculate the stellar mass and the helium core mass
of the primary star at the helium flash (hereafter
$M_{\rm HF}$ and $M_{\rm c,HF}$, respectively).
Then,  $M_{\rm HF}$, $M_{\rm c,HF}$, and a lifetime spent on the HB are used
to obtain the exact position of the primary star (e.g.,
effective temperature and luminosity)
on the HB in the Hertzsprung-Russell
(H-R) diagram by interpolating  among constructed HB evolutionary tracks.
All the HB evolutionary tracks are constructed using
modules for experiments in stellar astrophysics (MESA;
Paxton et al. 2011; see Paper I for details).
Finally, we transform the effective temperature and luminosity
of each HB star into $B-V$ colors and absolute magnitudes, $M_{\rm V}$,
to obtain the synthetic HB morphology in CMD.
All these processes are carried out
for different metallicities and ages to
investigate the effects of these two parameters on
the final results.

\subsection{Stellar evolution code } \label{bozomath}

Eggleton's stellar evolution code  was developed in the early 1970s
(Eggleton 1971, 1972, 1973).
This code has been updated with the latest input
physics over the past four decades (Han et al. 1994; Pols et al. 1995, 1998).
Now, this code is being applied extensively in the field of stellar evolution.
In our calculations, the ratio of mixing length to local pressure scale height, $\alpha=l/H_{\rm P}$, is set to 2.0.
The convective overshooting parameter, $\delta_{\rm ov}$, is set to 0.12 (Pols et al. 1997).
The opacity used in our calculations was compiled by Chen \& Tout (2007).
We obtain the initial hydrogen mass fraction, $X$,
by $X$ = 0.76-3$Z$, where $Z$ is the metallicity (Pols
et al. 1998).

\subsection{Tidally enhanced stellar wind } \label{bozomath}

We use the same equation as in Paper I to describe the
tidally enhanced stellar wind during binary evolution,
\begin{equation}
\dot{M}=-\eta4\times10^{-13}(RL/M)\{1+B_{\rm w}\times \rm min [\it(R/R_{\rm L})^{6}, \rm 1/2^{6}]\},
\end{equation}
where $\eta$ is the Reimers mass-loss efficiency,
$R_{L}$ is the radius of the Roche lobe, $B_{\rm w}$ is the tidal enhancement efficiency,
and $R$, $L$, $M$ are in solar units.
This form of stellar wind was first suggested by
Tout \& Eggleton (1988) to explain the mass inversion phenomenon
found in some RS CVn binary systems (see Paper I for
detail).
We incorporate equation (1) into Eggleton's stellar evolution
code to calculate $M_{\rm HF}$ and  $M_{\rm c,HF}$
of the primary star.

From our previous result in Paper I,
the HB morphology is not very sensitive to
the tidal enhancement parameter,$B_{\rm w}$.
For this reason,  $B_{\rm w}$ is set to 10000 (e.g.,
the typical value used in Tout \& Eggleton 1988) in all
sets of calculations in this paper.
Reimers mass-loss efficiency (Reimers 1975),
$\eta$, is set to 0.45 (see the discussion in Section~4.2).
To study the effects of
metallicity and age on
the HB morphology by considering
tidally enhanced stellar wind,
we carry out  two sets of calculations:
(i) for a fixed age of 13 Gyr,
we use three different metallicities, namely
$Z$=0.0001, 0.001, and 0.02. For these three metallicities,
the stellar mass of the primary
at zero-age main sequence (hereafter $M_{\rm ZAMS}$)
are 0.80, 0.83 and 0.98$M_{\odot}$ respectively.
(ii) For a fixed metallicity of GCs, $Z$=0.001, we use five
$M_{\rm ZAMS}$ in our model calculations, namely
0.80, 0.83, 0.85, 0.89, and 0.97$M_{\odot}$.
These five masses correspond to cluster ages of
14, 13, 12, 10, and 7 Gyr,  respectively.
The detailed information for
the two sets of calculations are
shown in Table 1. The columns from
left to right provide the metallicity,
$M_{\rm ZAMS}$, and
the age of the primary star at helium flash, respectively.
The results of the two sets of calculations are
given in Tables~2 and 3 (see Section 3.1).

The mass ratio of primary to secondary is
set to 1.6 (We
used different mass ratios of primary to secondary in
Paper I, but it has little influence on our results.)
We assume that the mass accreted by
the secondary from the stellar wind is retained in the binary system,
and the angular momentum that  leaves the binary system due
to the stellar wind  is attributed to the primary star.

\subsection{Initial binary samples} \label{bozomath}

To obtain the synthetic HB morphology in CMD,
we generate several  groups of binary systems,
in which the metallicity and $M_{\rm ZAMS}$ of the primary stars
correspond to
the values given in Table~1. The mass ratio of
primary to secondary is set to 1.6 (see
Section~2.2).
The initial orbital periods of all 
binary systems  are produced by
Monte Carlo simulations.
The distribution of
separation in binary is constant in $\log a$
($a$ is the separation) and falls off smoothly at small separations,
\begin{equation}
a\cdot n(a)=\left\{
 \begin{array}{lc}
 \alpha_{\rm sep}(a/a_{\rm 0})^{\rm m}, & a\leq a_{\rm 0},\\
\alpha_{\rm sep}, & a_{\rm 0}<a<a_{\rm 1},\\
\end{array}\right.
\end{equation}
where $\alpha_{\rm sep}\approx0.07$, $a_{\rm 0}=10\,R_{\odot}$,
$a_{\rm 1}=5.75\times 10^{\rm 6}\,R_{\odot}=0.13\,{\rm pc}$ and
$m\approx1.2$. This distribution implies that the number of wide
binaries per logarithmic interval are equal and about 50\% of
stellar systems have orbital periods less than 100\,yr (Han et al. 2003).

\section{Results}

\subsection{$M_{\rm HF}$ and $M_{\rm c,HF}$ of the primary star
for different metallicities and ages.}
\label{bozomath}

   \begin{table}
\renewcommand{\arraystretch}{1.2}

\centering

      \begin{minipage}[]{100mm}
   \caption[]{Information for the two sets of
   calculations in this paper.}
   \end{minipage}\\

    \begin{tabular}{c|ccc}
    \hline
      &$Z$  & $M_{\rm ZAMS} (M_{\odot})$  & age (Gyr) \\
     \hline
      &0.0001 &  0.80       & 13        \\
set(i)&0.001  &  0.83       & 13        \\
      &0.02   &  0.98       & 13        \\
      \hline\hline\noalign{\smallskip}
      &0.001 &  0.80       & 14       \\
      &0.001 &  0.83       & 13       \\
set(ii)&0.001 &  0.85       & 12       \\
      &0.001 &  0.89       & 10        \\
      &0.001 &  0.97       & 7       \\

              \hline\noalign{\smallskip}
   \end{tabular}
      \end{table}

Table~2 shows the results of our
model calculation for different
metallicites, and it gives the  $M_{\rm HF}$, $M_{\rm c,HF}$,  and
the envelope mass of the primary
stars at the helium flash ($M_{\rm env}$)
with various initial binary orbital periods
at a fixed age of 13 Gyr.
The results in Table~2 correspond to the calculation of set (i) (see
Section~2.2).  Recall that
$B_{\rm w}$ is set to $10^{4}$ and $\eta$ is set to 0.45.
The mass ratio of primary-to-secondary, $q$, is set to be 1.6.
The columns from
left to right  provide the
initial orbital periods, $M_{\rm HF}$,  $M_{\rm c,HF}$,
and $M_{\rm env}$, respectively.

For each metallicity, the first
binary orbital period in Table~2 is
a minimum period (e.g., log$P/\rm day=3.126$ for
$M_{\rm ZAMS}=0.8M_{\odot}$ at metallicity of $Z=0.0001$),
above which a helium flash could take place.
For orbital periods shorter than
this critical value\footnote
{If the binary orbital period is
short enough to make the the primary star
fill its Roche lobe, then a Roche lobe overflow (RLOF) or
a common envelope (CE) will begin in this binary system, but
this is beyond the scope of present work.},
the primary stars experience
too much mass loss on the RGB due to
the tidally enhanced stellar wind, and
fail to ignite helium in their cores.
Therefore, these primary stars will
not undergo the HB phase and
evolve into helium white dwarf (WD) cooling curve directly.
With the increase in binary
orbital period, the amount of
mass loss of the primary star on the RGB
decreases for each metallicity. When the
initial binary orbital period is
long enough, the tidally enhanced stellar
wind becomes unimportant and has little effect on the mass loss
of red giant primary (One can see that
the  $M_{\rm env}$ of primary stars with
log$P/\rm day=3.55$ and log$P/\rm day$=10.0
for $M_{\rm ZAMS}=0.8M_{\odot}$ at the metallicity
of $Z=0.0001$ are nearly the same, which means that
they experience nearly the same
amount of mass loss on the RGB.) At this time,
the primary star just loses its
envelope mass through Reimers mass-loss law (Reimers 1975).

Similar to Table~2, Table~3 presents our
model calculation results for
different ages at metallicity of $Z$=0.001,
which corresponds to the calculation of set (ii) (see Section~2.2).

  \begin{table}
\renewcommand{\arraystretch}{0.8}

\centering

      \begin{minipage}[]{80mm}
   \caption[]{$M_{\rm HF}$, $M_{\rm c,HF}$, and $M_{\rm env}$ of the primary
    with various initial binary orbital periods for
   different metallicities at a fixed age of 13 Gyr.
   Here, $B_{\rm w}=10000$, $\eta=0.45$, $q=1.6$.}
   \end{minipage}\\

    \begin{tabular}{cccc}

    \hline\noalign{\smallskip}

      log$P/\rm day$  & $M_{\rm HF}$($M_{\odot}$)
      & $M_{\rm c,HF}$($M_{\odot}$)  & $M_{\rm env}$($M_{\odot}$)\\
           \hline\noalign{\smallskip}

     $Z$=0.0001,  &$M_{\rm ZAMS}=0.80$ $M_{\odot}$, & age=13 Gyr     \\

      \hline\noalign{\smallskip}
      3.1260 & 0.4869 & 0.4854 & 0.0015   \\
      3.1500 & 0.4904 & 0.4890 & 0.0014   \\
      3.2000 & 0.4976 & 0.4964 & 0.0012   \\
      3.2500 & 0.5047 & 0.5018 & 0.0029   \\
      3.3000 & 0.5328 & 0.5024 & 0.0304   \\
      3.3500 & 0.5688 & 0.5022 & 0.0666   \\
      3.4000 & 0.5932 & 0.5021 & 0.0911   \\
      3.4500 & 0.6088 & 0.5020 & 0.1068   \\
      3.5000 & 0.6186 & 0.5019 & 0.1167   \\
      3.5500 & 0.6247 & 0.5018 & 0.1229   \\
      10.000 & 0.6351 & 0.5017 & 0.1334    \\
          \hline\hline\noalign{\smallskip}
      $Z$=0.001,  &$M_{\rm ZAMS}=0.83$ $M_{\odot}$, & age=13 Gyr     \\
            \hline\noalign{\smallskip}
      3.2590 & 0.4725 & 0.4714 & 0.0011    \\
      3.3000 & 0.4783 & 0.4774 & 0.0009    \\
      3.3500 & 0.4854 & 0.4845 & 0.0009    \\
      3.4000 & 0.4926 & 0.4895 & 0.0031    \\
      3.4500 & 0.5309 & 0.4893 & 0.0416    \\
      3.5000 & 0.5677 & 0.4892 & 0.0785    \\
      3.5500 & 0.5935 & 0.4891 & 0.1044    \\
      3.6000 & 0.6103 & 0.4890 & 0.1213    \\
      3.6500 & 0.6208 & 0.4889 & 0.1319    \\
      10.000 & 0.6386 & 0.4887 & 0.1499    \\
                \hline\hline\noalign{\smallskip}
      $Z$=0.02,  &$M_{\rm ZAMS}=0.98$ $M_\odot$, & age=13 Gyr     \\
            \hline\noalign{\smallskip}
      3.5500 & 0.4504 & 0.4496 & 0.0008    \\
      3.6000 & 0.4571 & 0.4563 & 0.0008    \\
      3.7000 & 0.4709 & 0.4699 & 0.0010    \\
      3.7500 & 0.5171 & 0.4703 & 0.0468    \\
      3.8000 & 0.5910 & 0.4703 & 0.1207    \\
      3.8500 & 0.6491 & 0.4702 & 0.1789    \\
      3.9000 & 0.6853 & 0.4701 & 0.2152    \\
      3.9500 & 0.7072 & 0.4700 & 0.2372    \\
      4.0000 & 0.7206 & 0.4699 & 0.2507     \\
      4.5000 & 0.7421 & 0.4698 & 0.2723    \\
      10.000 & 0.7423 & 0.4698 & 0.2725    \\
      \hline\hline

         \end{tabular}
      \end{table}

  \begin{table}
\renewcommand{\arraystretch}{0.8}

\centering

      \begin{minipage}[]{80mm}
   \caption[]{$M_{\rm HF}$, $M_{\rm c,HF}$, and $M_{\rm env}$ of the primary
    with various initial binary orbital periods
   for different ages at $Z=0.001$ .
   Here, $B_{\rm w}=10000$, $\eta=0.45$, $q=1.6$.}
   \end{minipage}\\

    \begin{tabular}{cccc}

    \hline\noalign{\smallskip}

      log$P/\rm day$  & $M_{\rm HF}$($M_{\odot}$)
      & $M_{\rm c,HF}$($M_{\odot}$)  & $M_{\rm env}$($M_{\odot}$)\\
           \hline\noalign{\smallskip}

     $Z$=0.001,  &$M_{\rm ZAMS}=0.80$ $M_{\odot}$, & age=14 Gyr     \\

      \hline\noalign{\smallskip}
      3.2980 & 0.4734 & 0.4724 & 0.0010    \\
      3.3600 & 0.4817 & 0.4808 & 0.0009    \\
      3.4200 & 0.4894 & 0.4884 & 0.0010    \\
      3.4500 & 0.4933 & 0.4902 & 0.0031    \\
      3.5000 & 0.5210 & 0.4900 & 0.0310    \\
      3.5500 & 0.5463 & 0.4899 & 0.0564    \\
      3.6000 & 0.5641 & 0.4898 & 0.0743    \\
      3.6500 & 0.5758 & 0.4897 & 0.0861    \\
      10.000 & 0.5959 & 0.4896 & 0.1063    \\
          \hline\hline\noalign{\smallskip}
      $Z$=0.001,  &$M_{\rm ZAMS}=0.83$ $M_{\odot}$, & age=13 Gyr     \\
            \hline\noalign{\smallskip}
      3.2590 & 0.4725 & 0.4714 & 0.0011    \\
      3.3000 & 0.4783 & 0.4774 & 0.0009    \\
      3.3500 & 0.4854 & 0.4845 & 0.0009    \\
      3.4000 & 0.4926 & 0.4895 & 0.0031    \\
      3.4500 & 0.5309 & 0.4893 & 0.0416    \\
      3.5000 & 0.5677 & 0.4892 & 0.0785    \\
      3.5500 & 0.5935 & 0.4891 & 0.1044    \\
      3.6000 & 0.6103 & 0.4890 & 0.1213    \\
      3.6500 & 0.6208 & 0.4889 & 0.1319    \\
      10.000 & 0.6386 & 0.4887 & 0.1499    \\
                \hline\hline\noalign{\smallskip}
      $Z$=0.001,  &$M_{\rm ZAMS}=0.85$ $M_\odot$, & age=12 Gyr     \\
            \hline\noalign{\smallskip}
      3.2360 & 0.4719 & 0.4708 & 0.0011    \\
      3.3000 & 0.4813 & 0.4803 & 0.0010    \\
      3.3500 & 0.4886 & 0.4875 & 0.0011    \\
      3.4000 & 0.5146 & 0.4889 & 0.0257    \\
      3.4500 & 0.5637 & 0.4889 & 0.0748    \\
      3.5000 & 0.6008 & 0.4887 & 0.1121    \\
      3.5500 & 0.6251 & 0.4886 & 0.1365    \\
      3.6000 & 0.6405 & 0.4884 & 0.1521    \\
      3.6500 & 0.6502 & 0.4883 & 0.1619    \\
      10.000 & 0.6662 & 0.4883 & 0.1779    \\
      \hline\hline

    $Z$=0.001, &$M_{\rm ZAMS}=0.89$ $M_\odot$, & age =10 Gyr   \\

      \hline\noalign{\smallskip}
      3.1950 & 0.4708 & 0.4698 & 0.0010    \\
      3.2500 & 0.4792 & 0.4793 & 0.0009    \\
      3.3000 & 0.4869 & 0.4859 & 0.0010    \\
      3.3500 & 0.5162 & 0.4879 & 0.0283    \\
      3.4000 & 0.5832 & 0.4879 & 0.0953    \\
      3.4500 & 0.6338 & 0.4877 & 0.1461    \\
      3.5000 & 0.6665 & 0.4876 & 0.1789    \\
      3.5500 & 0.6868 & 0.4875 & 0.1993    \\
      3.6000 & 0.6993 & 0.4874 & 0.2119    \\
      3.6500 & 0.7071 & 0.4874 & 0.2197    \\
      10.000 & 0.7201 & 0.4873 & 0.2328    \\

      \hline\hline\noalign{\smallskip}
        $Z$=0.001, &$M_{\rm ZAMS}=0.97$ $M_\odot$, & age =7 Gyr     \\
      \hline\noalign{\smallskip}
      3.1250 & 0.4687 & 0.4677 & 0.0010    \\
      3.2000 & 0.4808 & 0.4799 & 0.0009    \\
      3.2500 & 0.4894 & 0.4861 & 0.0033    \\
      3.3000 & 0.5827 & 0.4862 & 0.0965    \\
      3.3500 & 0.6741 & 0.4861 & 0.1880    \\
      3.4000 & 0.7319 & 0.4860 & 0.2459    \\
      3.4500 & 0.7669 & 0.4859 & 0.2810    \\
      3.5000 & 0.7882 & 0.4858 & 0.3024    \\
      3.5500 & 0.8013 & 0.4858 & 0.3155    \\
      3.6000 & 0.8094 & 0.4857 & 0.3237    \\
      10.000 & 0.8227 & 0.4856 & 0.3371    \\
      \hline\hline

         \end{tabular}
      \end{table}

      \begin{figure}
\centering
\includegraphics[width=100mm,angle=0]{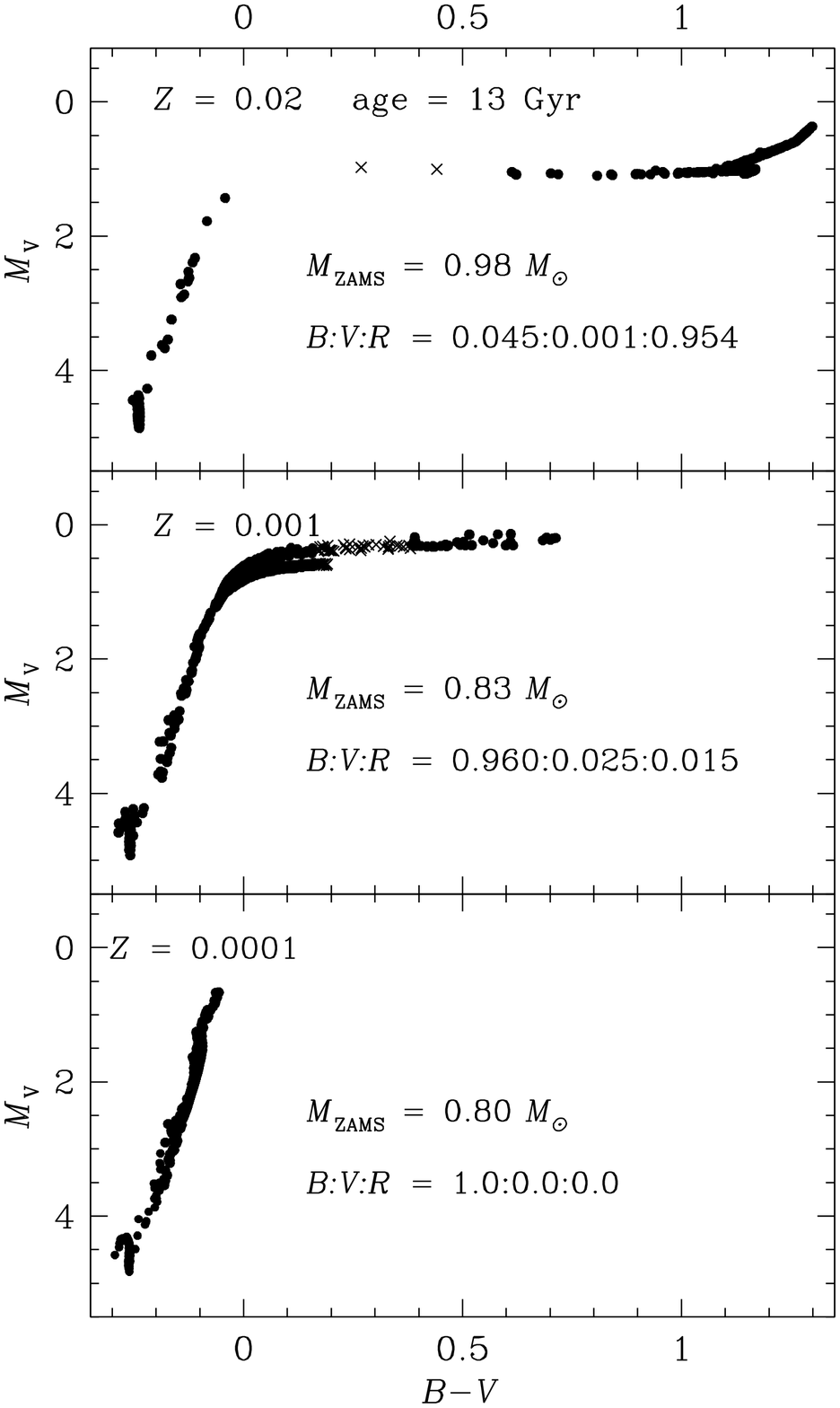}
\begin{minipage}[]{90mm}
\caption{Synthetic HB morphology produced by tidally enhanced stellar
wind for different metallicities at a fixed age
of 13 Gyr. The metallicities from
the top panel to the bottom panel are $Z$=0.02, 0.001, 0.0001,
for which the primary mass are 0.98, 0.83, and 0.80$M_\odot$,
respectively. HB stars in RR Lyrae strip (defined by
the vertical region of $3.80\leq$ log$T_{\rm eff}$ $\leq3.875$
in the H-R diagram, Koopmann et al. 1994; Lee et al. 1990) are denoted by
crosses. Other HB stars are denoted by solid dots. The label of
$B:V:R$ in this figure is the number ratio of HB stars
located in the region bluer than (or to the left of), within,  and
redder than (or to the right of) the RR Lyrae instability strip in the
H-R diagram.}. \end{minipage}

   \label{Fig1}
\end{figure}

\begin{figure}
\centering
\includegraphics[width=100mm,angle=0]{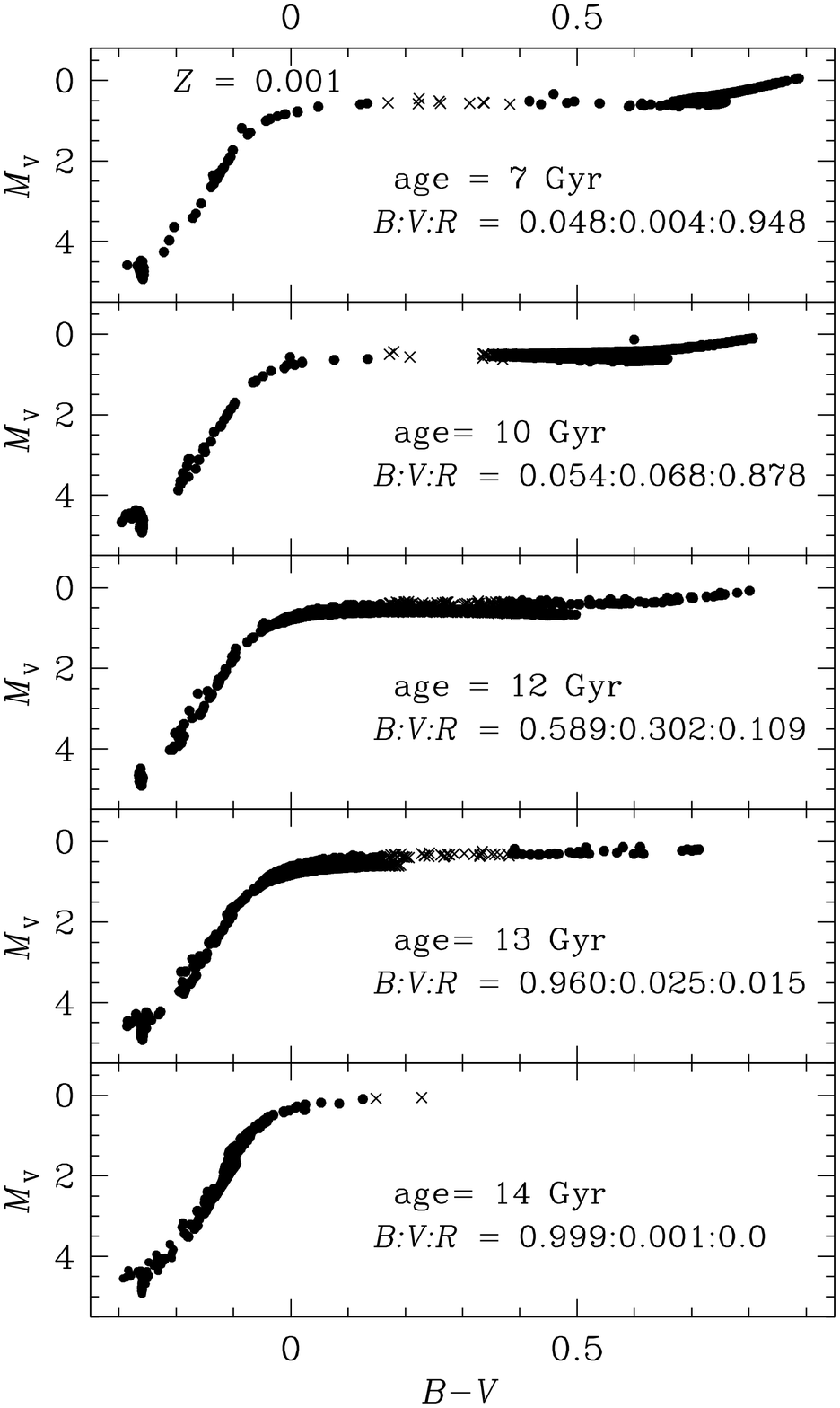}
\begin{minipage}[]{90mm}
\caption{Similar to Fig.1, but for
different ages at $Z$=0.001. The $M_{\rm ZAMS}$ of
the primary star from the
top to bottom panels are 0.97, 0.89, 0.85, 0.83, and 0.80$M_\odot$,
which corresponds to
cluster ages of 7, 10, 12, 13, and 14 Gyr, respectively.   } \end{minipage}
   \label{Fig2}
\end{figure}

\subsection{Synthetic HB morphology in CMD for different metallicities}

To obtain the position of
each sample star on the HB in the H-R diagram
(e.g., effective temperatures and luminosities),
we need to know  $M_{\rm HF}$,  $M_{\rm c,HF}$, and the time
spent on HB phase which means how long the HB star has been
evolved from zero-age HB (ZAHB).
The value of $M_{\rm HF}$ and $M_{\rm c,HF}$
for each primary star in this set of calculation
are obtained by interpolating with
the results presented in Table~2.
The time spent on the HB is generated by a uniform
random number between 0 and the HB lifetime, $\tau
_{\rm HB}$ (Rood 1973; Lee et al. 1990; Dalessandro et al. 2011).   Here,
$\tau_{\rm HB}$ is set to be the lifetime of  HB star
with the lowest stellar mass
among the constructed HB evolutionary tracks,
which means
that this star has the longest lifetime on the HB.  Therefore,
some of HB stars are given a time longer than their
lifetimes on the HB, and these stars are considered to evolve
into the next evolutionary phase (e.g., AGB or WD).
This process is equivalent to the scenario that stars enter the HB
from the RGB at a constant rate (see Lee et al. 1990).
We use $M_{\rm HF}$,  $M_{\rm c,HF}$, and the time spent
on HB phase to obtain the exact position of
each primary star on HB in the H-R diagram by
interpolating among constructed HB evolutionary tracks.
After that, we transform the temperature and luminosity of each HB star into
$B-V$ color and absolute magnitude, $M_{\rm V}$,
using the Basel stellar spectra library
(Lejeune et al. 1997, 1998).
Finally, we obtain the synthetic HB in CMD.

Figure.1 shows the synthetic HB for three different metallicities
under the tidally enhanced stellar wind.
The metallicities from the top to bottom
panels in Fig.1 are $Z$=0.02, 0.001, and 0.0001, respectively.
The corresponding $M_{\rm ZAMS}$  are
0.98, 0.83, and 0.80$M_\odot$.
In the top panel of Fig.1, the synthetic HB has
a high metallicity of $Z$=0.02. Most of
HB stars in this panel locate in red HB (e.g.,
95\%), while stars in blue and extreme
HB are just about 5\%. One can see clearly in this panel that
there is a distinct gap between red and blue HB, and a few of
HB stars are located in the RR Lyrae instability strip.
However, this bimodal HB morphology is not
presented clearly in the other two panels of Fig.1.
We discuss this result in Section~4.1.
For the whole figure, with the metallicity decreasing from the
top to bottom panels,
more and more HB stars settle on blue and
extreme HB. For the extreme case, in the bottom panel of
Fig.1, which has the lowest metallicity of
$Z$=0.0001 in our calculations,  all HB stars are located 
in the region that is bluer than the RR Lyrae instability strip, with
no red HB stars and RR Lyrae stars are produced.
Figure.1 shows the typical effect
of metallicity (the first parameter) on HB morphology in
GCs, which is that metal-poor GCs present
bluer HB morphology than the metal-poor ones.
This is because the metal-poor GCs have a lower stellar mass
than the metal-rich ones at the tip of RGB for a fixed cluster age.
As a result, assuming that
the mass-loss law on the RGB is the same,
it is much easier for the metal-poor stars
to settle on blue HB positions than for the metal-rich ones.
Moreover, for the same envelope mass, the
metal-poor HB stars have much higher
effective temperatures on the ZAHB than
the metal-rich ones owing
their lower opacity in envelopes. This fact could also
make the metal-poor HB stars occupy
blue HB positions more easily than
the metal-rich ones.

\subsection{Synthetic HB morphology in CMD for different ages}

The synthetic HB morphology in CMD for different ages at a fixed
metallicity of $Z=0.001$
are obtained in a similar way to the one used in Section~3.2, but
the value of  $M_{\rm HF}$ and $M_{\rm c,HF}$
for each sample star
in this set of calculation
are obtained by interpolating
with the results in Table~3.

Figure.2 shows five synthetic HB in CMD with different ages of GCs
at metallicity of $Z$=0.001.
The  $M_{\rm ZAMS}$ of the primary star in Fig.2 from the top to
bottom panels are 0.97, 0.89, 0.85, 0.83, and 0.80$M_\odot$.
These five masses correspond to
cluster ages of about 7, 10, 12, 13, and 14 Gyr, respectively.
The label of $B:V:R$ in Fig.2 has the same meaning as
in Fig.1.

One can see clearly in Fig.2 that,  with the age
increasing  from the top to the bottom
panel, more and more HB stars are located in blue and
extreme HB. Especially in bottom panel of
Fig.2, which corresponds to the largest cluster age of about 14
Gyr in our model calculations, the synthetic HB is nearly a whole blue HB,
and no red HB stars are produced.  Figure.2 indicates that old GCs will present
bluer HB morphology than the young ones. This is because that
old GCs have a  lower
stellar mass than the young ones at the RGB tip for a fixed metallicity.
Therefore, the stars in
old GCs settle more easily  on
blue HB positions than the stars in young GCs if
they follow the same mass-loss law on the RGB.

\section{Discussion}

\begin{figure}
\centering
\includegraphics[width=100mm,angle=0]{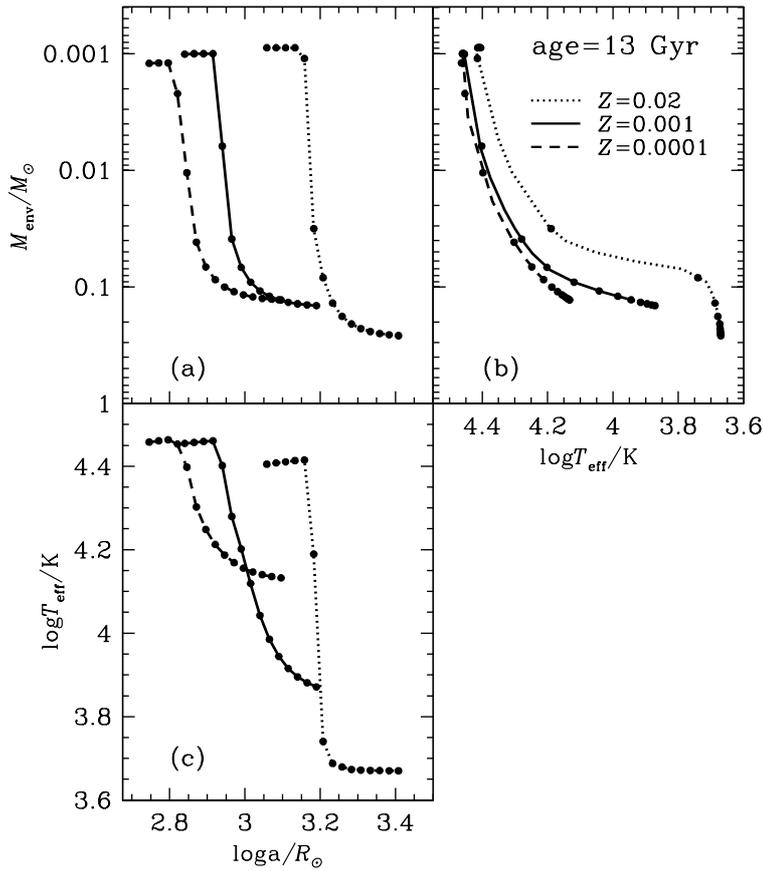}
\begin{minipage}[]{90mm}
\caption{Panel (a) gives the envelope mass distribution of
ZAHB stars produced by the tidally enhanced stellar wind.
Panel (b) shows the relationship between the envelope mass
and effective temperature of ZAHB stars. Panel (c)
presents the effective temperature distribution of ZAHB stars produced
by the tidally enhanced stellar wind.  The dashed, solid, and
dotted curves in each panel correspond to
the model calculations for three different metallicities as in Fig.1,
namely $Z$ =0.0001, 0.001, 0.02, respectively.  The two
adjacent solid dots in each curve of these three panels denote
a fixed separation interval of  $\Delta \log a/R_\odot$=0.025 ($a$ is
the separation of binary system). } \end{minipage}
   \label{Fig3}
\end{figure}

\subsection{Bimodality on HB }

From the top panel of Fig.1, in which the metallicity is $Z$=0.02,
one can see that there is
a distinct gap between red and blue HB.
However, this bimodal HB distribution is not seen clearly in our synthetic
HB for $Z$=0.001 and 0.0001 in Fig.1, which indicates that
metal-rich GCs are more likely to
form a bimodal HB than the metal-poor ones.
To discuss the physical reason for this result,
we present three panels in Fig.3.
The dashed, solid and dotted curves in each
panel correspond to the model calculations
for three different metallicities as in Fig.1, namly
$Z$=0.0001, 0.001, and 0.02. 

Panel (a) gives the
envelope mass distribution of ZAHB stars produced by
the tidally enhanced stellar wind.
Since the distribution of separation in $\log a$
is constant when the separation, $a$, is larger than
10 $R_\odot$ (see Section~2.3), the number of the initial binary systems
in each fixed separation interval
(i.e., 0.025) is
the same. Therefore, the higher the concentration of the solid dots in
each curve, the more ZAHB stars will be produced
in this envelope mass range.
One can see clearly from panel (a) of Fig.3
that the envelope mass distributions produced
by the tidally enhanced stellar wind
for three different metallicities
are similar and that the number of ZAHB stars
with their envelope masses in  the rough range of
$0.003 M_{\odot} \lesssim M_{\rm env} \lesssim 0.06M_{\odot}$
is obviously less than the number of ZAHB stars with
the envelope mass beyond this range.

Panel (b) of Fig.3 shows the relationship between
the envelope mass and the effective temperatures of
ZAHB stars. One can see clearly that, for $Z=0.02$,
a large gap is presented within the effective temperature range
of $3.8\lesssim \rm {log}T_{\rm eff} \lesssim 4.3$.
However, for $Z=$ 0.001 and 0.0001, the sparse area
on ZAHB occupies a much narrower and bluer effective
temperature range than that of $Z=0.02$. This is
because the opacity of metal-poor HB stars in their
envelopes is less than the metal-rich ones.
Furthermore, due to the decreasing sensitivity of
$B-V$ color to effective temperature towards higher
temperatures (Moehler 2010) and due to the vertical evolution of hot HB stars
in H-R diagram (see Fig.1 in Paper I), the small gap on HB
for metal-poor GCs in CMD may become obscure.

Panel (c) in Fig.3 is the combination of panels (a) and (b), which
shows the relationship between the initial separations of binary systems and
the effective temperatures of ZAHB stars produced by
tidally enhanced stellar wind.
One can see that the distribution of HB stars
is clearly bimodal for $Z=0.02$.
However, this feature becomes more and more obscure
with decreasing metallicity, which
means that in our model calculation, metal-rich
GCs are more likely to form a bimodal HB morphology than
the metal-poor ones.

\subsection{age-HBR diagram in GCs}

\begin{figure}
\centering
\includegraphics[width=90mm,angle=0]{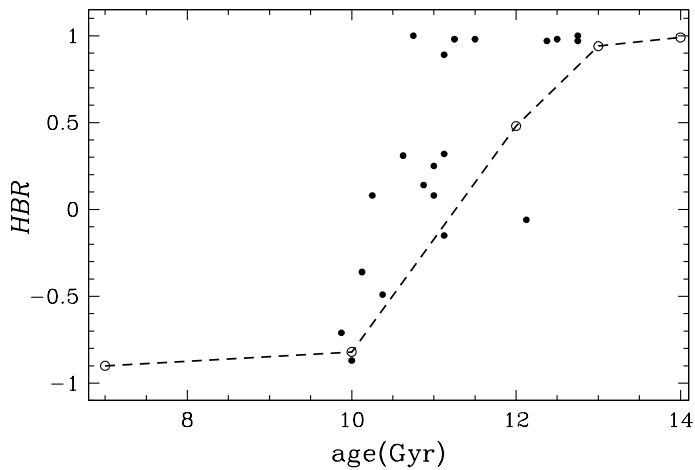}
\begin{minipage}[]{90mm}
\caption{The relationship between the age and the HB morphology parameter,
$HBR$,  in GCs. The solid circles are the selected
GCs in our Galaxy, which are in the metallicity range of
-1.6$<$[Fe/H]$<$-1.1. The open circles denote our model calculations
in Fig.2, which from left to right correspond to
cluster ages of 7, 10, 12, 13, and 14 Gyr, respectively. } \end{minipage}
   \label{Fig4}
\end{figure}

One can see from Fig.2 that old GCs present
bluer HB morphology than the young ones.
To compare this result with the observation in GCs,
we show in Fig.4 the relationship
between the age of GCs in our Galaxy and
the parameter, $HBR$, which is used to describe the
HB morphology of GCs (Lee et al. 1990; Lee et al. 1994).
The parameter, $HBR$, is defined as,
\begin{equation}
HBR=(B-R)/(B+V+R),
\end{equation}
where $B, V$, and $R$ have the same meanings as described in
Section~3.2. The value of the parameter, $HBR$,
is in the range of -1 to 1.
The value of -1 means that
all HB stars settle on red HB;
on the other hand, the value of 1
means that GC presents a whole blue HB.
Therefore, the larger the $HBR$ parameter,
the bluer the HB morphology in GCs.
The values of $HBR$ parameter for GCs in Fig.4  are
from the catalog  of Harris (1996),  and
the age of GCs used in this figure is from
Gratton et al. (2010)\footnote{The relative
ages of GCs in Gratton et al. (2010) are from Marin-Franch
et al. (2009) and De Angeli et al. (2005). }.
To compare the observation with our results,
we chose the GCs in the metallicity
range of $-1.6<\rm [Fe/H]<-1.1$ in Fig.4, which is around the
metallicity used in Fig.2 (i.e., $Z=0.001$ or [Fe/H] = -1.3).
This limit weakens the effect of metallicity
on HB morphology, thus the effect of age on
HB morphology could be revealed more clearly.

One can see in Fig.4 that,  for the selected
GCs in our Galaxy, the value of $HBR$ increases
with the increasing age of the GCs, which means
the HB morphology becomes bluer when the ages of
the GCs become older. This result
is consistent with the one obtained from
Fig.2, and also can be seen clearly from the dashed line in Fig.4.
However, for a fixed age of GCs, our synthetic HB produced
by a tidally enhanced stellar wind in Fig.4 is a little redder than
the observed ones. This may be because 
we do not consider the common envelope (CE) and Roche lobe overflow (RLOF)
processes in binary evolution
for our calculations (see Section 3.1 and Paper I).
These processes can produce EHB stars ( Han et al. 2002, 2003), and
thus make HB bluer.

We also adopted a different value of Reimers mass-loss efficiency
(e.g., $\eta$=0.25, which is the value we used in Paper I)
in our model calculations.
We found that, for $\eta$=0.25, the result is similar
to the one in Fig.4.  However, to produce a
blue HB morphology (e.g., $HBR \gtrsim 0.5$), the age of GCs should be
greater than 15 Gyr. This indicates that
$\eta$=0.25 is too small to produce the observed HB morphology in
GCs. Renzini \& Fusi Pecci (1988) demonstrated that observed HB morphologies
in GCs with $Z \simeq$ 0.001 demand $\eta=0.4\pm0.04$. That is why
we use the value of $\eta=0.45$ in this paper, but it does not
influence the results we obtained in Paper I, since we can
obtain the same results using a little younger age of GCs
for a higher value of $\eta$.

\subsection{Effects of binary fraction on HB morphology}

      \begin{figure}
\centering
\includegraphics[width=100mm,angle=0]{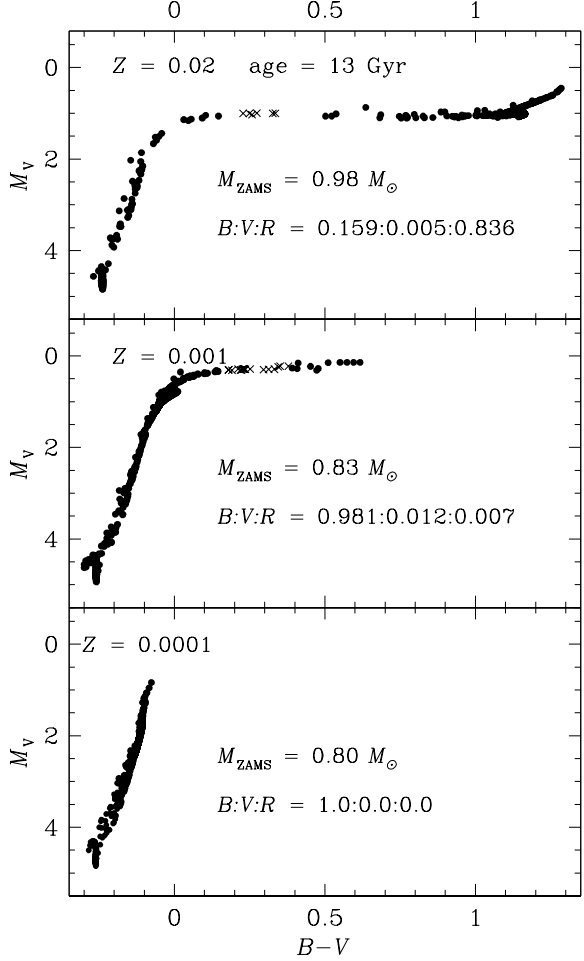}
\begin{minipage}[]{90mm}
\caption{Same as Fig.1, but with a higher binary fraction of 80\% rather
than 50\%.}. \end{minipage}

   \label{Fig5}
\end{figure}

      \begin{figure}
\centering
\includegraphics[width=100mm,angle=0]{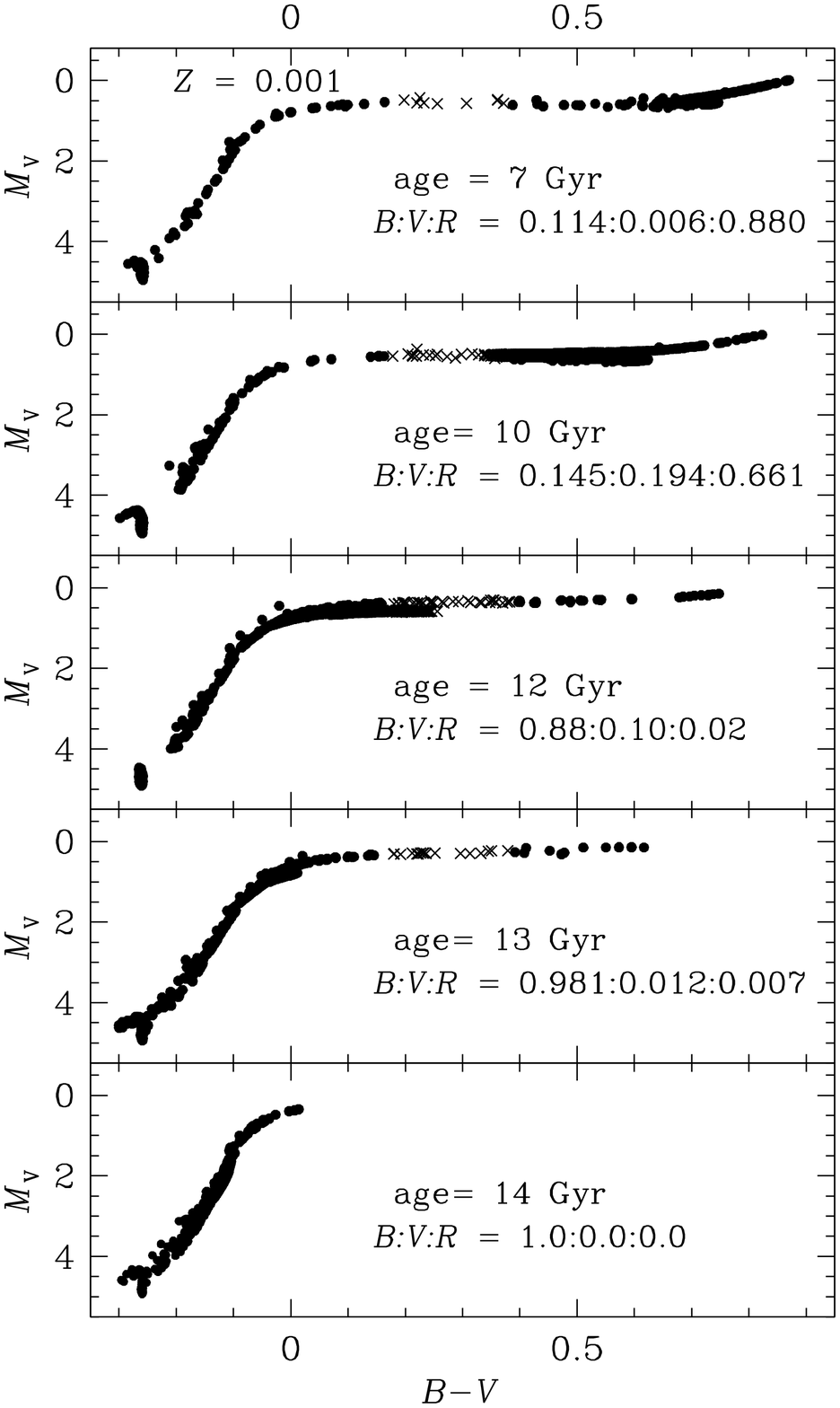}
\begin{minipage}[]{90mm}
\caption{Same as Fig.2, but with a higher binary fraction of 80\% rather
than 50\%.}. \end{minipage}

   \label{Fig6}
\end{figure}

In our model calculations, all
stars are in binary systems, but the binaries with long
orbital periods are actually single stars.
The separation distribution of binary systems described in Section~2.3
implies that about 50\% of binary systems have orbital periods
that all shorter than 100 yr.
To investigate the effects of binary fractions on our final results,
we also adopted various binary fractions in our calculations
(i.e., 10\%, 15\%, 20\%, 30\%, and 80\%)\footnote{All these binary fractions
 are for the binary systems with their orbital periods
less than 100 yr.}.
The result of model calculation for a binary fraction of 80\% is shown in
Figs.5 and 6. The input parameters in Figs.5
and 6 are the same as in Figs.1 and 2, except for
the binary fraction.

By comparing Figs.5 and 1, one can see that,
with the binary fraction increasing, the number ratio of HB stars
bluer than the RR Lyrae instability strip increases,
while the number ratio of red HB stars decreases. 
For example, in the top panel of Fig.5, the number ratio of HB stars
bluer than RR Lyrae instability strip is 15.9\%, as opposed to 4.5\% in the top
panel of Fig.1; while the number ratio of red HB stars is 83.6\%, versus 95.4\%
in the top panel of Fig.1.
We can also obtain this result by comparing Figs.6 and 2, and  the
model calculations for binary fraction of 10\%, 15\%, 20\%, and 30\% show
similar results. 
These results indicate that higher binary fraction may make 
HB morphology become bluer. 
This is because the orbital period in which
tidally enhanced stellar wind significantly affects mass loss
of the primary star is from log$P$/day$\approx 3.1$
to log$P$/day$\approx 3.5$ (see Tables~2 and 3), and
the orbital periods of binaries in this range are less than 100 yr. Therefore, for a given
total number of binary samples, when the fraction
of binary systems with their binary orbital
periods less than 100 yr increases, more primary stars
will be influenced by tidally enhanced stellar winds, and then
more blue and extreme HB stars will be produced.

Our results indicate that higher binary fraction 
may make the HB morphology become bluer,
which implies that binary population
is a possible second parameter candidate.
However, Milone et al. (2012) estimate the binary fraction for 59 Galactic GCs
by analyzing the number of stars located on the
red side of the main-sequence fiducial line (also see
Sollima et al. 2007). Moreover, they
studied the relationship between the binary fraction
and the HB morphology parameters, and conclude that
there is weak or null impact of  binary populations
on HB morphology (see Fig.47 and Section~5.6.3 in their paper).
At first glance,
their results contradict ours. However,
the sample GCs in Milone et al. (2012) have different
metallicities and ages, which are very important
parameters that affect HB morphology in GCs (Gratton et al. 2010;
Dotter et al. 2010). Therefore,
Milone et al. (2012) did not remove the effects of metallicity and age on
HB morphology when studying the effect of binary fraction on
HB morphology, and this will significantly influence the
final results\footnote{For example, we used a binary fraction of 10\% in the 
calculation and found that, though the binary fraction is relatively low,
the synthetic HB for $Z$=0.0001 at 13 Gyr is still a pure blue HB (e.g., $HBR$=1).
It means that the effect of metallicity may mask the effect of binary 
fraction on HB morphology at very low metallicity.}.
On the other hand, our result that higher 
binary fraction may make HB morphology become bluer is obtained by comparing
GCs with fixed metalicity and age but different binary fractions.
This means that we removed the effects of metalicity and age when
studying the effects of binary fraction on HB morphology.
Thus, the results obtained by Milone et al. (2012)
cannot exclude  binary populations as a second
parameter candidate that may  affect HB morphology in GCs.

One can also see that, though the binary fraction is higher than in Fig.1,
the synthetic HB for $Z$=0.02 in the top panel of
Fig.5 still presents a distinct gap between red and blue HB, with
very few of the HB stars located in RR Lyrae instability strip, 
(Similar results are obtained for other different binary fractions.) 
This result indicates that binary fraction may have little effect on
the bimodality of HB, and the bimodality of HB may mainly depend on the metallicity of
GCs (see our discussion in Section~4.1).

\section{Conclusions}

In this paper, we considered 
tidally enhanced stellar winds during binary evolution to study the
effects on the HB morphology of GCs with different metallicities
and ages.
We found that metal-poor GCs should 
present bluer HB morphology than the metal-rich
ones, and this is consistent with previous work.
Furthermore, we found in our calculations
that the envelope-mass distributions of ZAHB stars
produced by a tidally enhanced stellar wind are very similar
for different metallicities. However,
the synthetic HB for $Z$=0.02 produced by tidally enhanced stellar
wind presents a distinct gap between red and blue
HB, while this feature is not seen clearly in the synthetic
HB for $Z$=0.001 and 0.0001.
We also found that old
GCs present bluer HB than the young ones, which is
also consistent with previous work. We compared our results
with the observation
in the age-$HBR$ diagram of GCs. Furthermore,
we studied the effect of binary fraction on
our final results, and found that higher binary fraction may
make HB morphology become bluer. We finally discussed our
results with recent observations.

\begin{acknowledgements}
We thank the anonymous referee for  valuable comments
and suggestions that
helped us to improve the paper.
This work is supported by the National
Natural Science Foundation of
China (Grant Nos.11033008, 11273053 and 11203065) 
and the Chinese Academy of Sciences (Grant No. KJCX2-YW-T24).

\end{acknowledgements}

\clearpage

\end{document}